\documentstyle[12pt]{article}
\begin{document}
\input amssym.tex

\title{External symmetry in general relativity}

\author{Ion I. Cot\u aescu\\ {\it The West University of Timi\c soara,}\\
{\it V. Parvan Ave. 4, RO-1900 Timi\c soara}}

\maketitle

\begin{abstract}
We propose  a generalization of the isometry transformations to the geometric 
context of the field theories with spin  where the local frames are explicitly 
involved. We define the external symmetry transformations as isometries 
combined with suitable tetrad gauge transformations and we show that these 
form a group which is locally isomorphic with the isometry one. We  point out 
that the symmetry transformations that leave invariant the equations of the 
fields with spin have generators with specific spin terms which represent new 
physical observables. The examples we present are the generators of the 
central symmetry and those of the maximal symmetries of the de Sitter and 
anti-de Sitter spacetimes derived in different tetrad gauge fixings.  

Pacs: 04.20.Cv, 04.62..+v, 11.30.-j
\end{abstract}
\   

\section{Introduction}
\

In general relativity \cite{SW,MTW,WALD} the development of the quantum field 
theory in curved spacetimes \cite{BD} give rise to many difficult problems 
related to the physical interpretation of the one-particle quantum modes that 
may indicate how to quantize the field. This is because the form and the 
properties of the particular solutions of the free field equations, in the 
cases when these can be analytically solved \cite{SOL1,SOL2}, are strongly 
dependent on the procedure of separation of variables and, implicitly, on the 
choice of the local chart. Moreover, when the fields  have spin the situation 
is more complicated since then the field equations and, therefore, the form of 
their particular solutions depend, in addition, on the tetrad gauge in which 
one works \cite{UK,SW}. In these conditions it would be helpful to use the 
traditional method of the quantum theory in flat spacetime based on the 
complete sets of commuting operators that determine the quantum modes as 
common eigenstates and give physical meaning to the constants of the 
separation of variables which are just the eigenvalues of these operators. A 
good step in this direction could be to proceed like in special relativity 
looking for the generators of the geometric symmetries similar to the 
familiar momentum, angular momentum and spin operators of the Poincar\' e 
covariant field theories \cite{W}. 
  
However, the relativistic covariance  in the sense of general relativity is 
too 
general to play the same role as the Lorentz or Poincar\' e covariance in 
special relativity. In other respects, the tetrad gauge invariance of the 
theories with spin represents another kind of general symmetry that is not 
able to produce itself conserved observables \cite{SW}. For this reason we 
have to concentrate only upon some special transformations which should form 
a well-defined Lie group with significant parameterization from the geometric 
point of view. Obviously, these must be just the isometry transformations 
that point out the symmetry of the background giving us the specific Killing 
vectors \cite{SW,WALD,ON}. The physical fields take over this symmetry  
transforming according to appropriate representations of the isometry group. 
In the case of the scalar vector or tensor fields these representations are 
completely defined by the well-known transformations rules under coordinate 
transformations since the isometries are in fact particular automorphysms. 
It remains open the problem of the behavior under isometries of the fields 
with half integer spin which explicitly depend on the tetrad gauge fixing. 
Another important  problem is how to define the generators of these 
representations for any spin. It is known that there is a standard 
operator-valued representation of the isometry group in the space of scalar 
functions whose  generators can be written with the help of the Killing 
vectors in a similar manner as the orbital angular momentum operators of 
special relativity. Then it is natural to ask how could be defined  
the corresponding  spin parts of the generators of the  
representations according which the fields with spin may transform.     

Our aim here is to propose a way to solve these problems. We start with the 
idea that if we intend to study the symmetry of a physical theory we must 
take into consideration the whole geometric context, including the positions 
of the local frames given by the tetrad fields, since the spin 
is measured just with respect to the axes of these frames. Therefore,   
the symmetry transformations must preserve not only the form of the metric 
tensor but the tetrad gauge too. These have to be isometries 
combined with suitable tetrad gauge transformations in such a manner to leave 
invariant the tetrad field components. Thus we define the {\em external 
symmetry} group and we show that this is locally isomorphic with the isometry 
group, having the same structure constants. Moreover, there are arguments 
that in fact this is isomorphic with the universal covering group of the 
isometry one. 

The next step is to 
define the operator-valued representations of the external symmetry group 
carried by spaces of fields with spin. We point out that these are induced  
by the linear finite-dimensional representations of the $SL(2,C)$ group.  
This is the motive why the symmetry transformations  which leave 
{\em invariant} the field equations have generators with a composite 
structure. 
These have the usual orbital terms of the scalar representation 
and, in addition, specific spin terms which depend on the choice of the 
tetrad gauge even in the case of the fields with integer spin.  In general, 
the spin and orbital terms do not commute to each other  apart of some 
special gauge fixings where the fields transform manifestly covariant under 
external symmetry transformations.            

These general results allow us to study two important examples, namely the 
central symmetry and the maximal symmetry of the de Sitter (dS) and 
anti-de Sitter (AdS) spacetimes. In the case of the central geometries we use 
central charts with Cartesian coordinates and the Cartesian tetrad gauge which 
allowed us recently to find new analytical solutions of the Dirac equation 
\cite{COTA}. We show that in this gauge fixing the central symmetry becomes 
global and, consequently, the spin terms of its generators are the same as 
those of special relativity \cite{BJDR,TH}. This is important from the 
technical point of view since in the largely used  diagonal tetrad gauge in 
spherical coordinates \cite{BW,D} we obtain that the spin terms are 
partially hidden.  For the dS and AdS spacetimes we calculate the  generators 
of the representations of the external symmetry group in central charts with 
our Cartesian gauge and in Minkowskian charts \cite{SW} with another gauge 
where the fields behave manifestly covariant under the Lorentz symmetry 
\cite{POL}.  

We start in the second section with a brief review of the relativistic 
covariance and gauge symmetry which will be treated together introducing 
the group of the combined transformations defined as gauge transformations 
followed by authomorphysms. The next section is devoted to our 
approach. Therein we define the external symmetry transformations, we 
show that these form a group and we study 
the operator-valued representations of this group and its Lie algebra. 
In Sec.4 and 5 we discuss the mentioned examples.    
 
We present our proposal at the level of the relativistic quantum mechanics 
in the sense of general relativity avoiding to consider the specific problems 
of the quantum field theory or to use too complicated mathematical methods. 
We work in natural units with $\hbar=c=1$.

\section{Relativistic covariance} 
\

In the Lagrangian field theory in curved spacetimes the relativistic 
covariant equations of scalar, vector or tensor fields arise from actions 
that are invariant under general coordinate transformations. Moreover, when 
the fields have spin in the sense of the $SL(2,C)$ symmetry then the action 
must be invariant under tetrad gauge transformations \cite{UK}. The first 
step to our approach we propose here is to  embed both these kind of 
transformations into new ones, called combined transformations, that will 
help us to understand the relativistic covariance in its most general terms.

\subsection{Gauge transformations}
\

Let us consider the curved spacetime $M$ and a local chart (natural frame)  
of coordinates $x^{\mu}, \mu=0,1,2,3$. Given a gauge, we denote by 
$e_{\hat\mu}(x)$ the tetrad fields that define the local frames, in each 
point $x$, and by $\hat e^{\hat\mu}(x)$ those 
of the corresponding coframes. These have the usual orthonormalization 
properties   
\begin{equation}\label{(duale)}
\hat e^{\hat\mu}_{\alpha}\, e_{\hat\nu}^{\alpha}=\delta^{\hat\mu}_{\hat\nu}
\,,\quad 
\hat e^{\hat\mu}_{\alpha}\, e_{\hat\mu}^{\beta}=\delta^{\beta}_{\alpha}
\,,\quad 
e_{\hat\mu}\cdot e_{\hat\nu}=\eta_{\hat\mu \hat\nu}\,, \quad
\hat e^{\hat\mu}\cdot \hat e^{\hat\nu}=\eta^{\hat\mu \hat\nu}\,, 
\end{equation}
where $\eta=$diag$(1,-1,-1,-1)$ is the Minkowki metric. 
From the line element
\begin{equation}\label{(met)}
ds^{2}=\eta_{\hat\mu \hat\nu}d\hat x^{\hat\mu}d\hat x^{\hat\nu}=
g_{\mu \nu}(x)dx^{\mu}dx^{\nu}\,,
\end{equation}   
expressed in terms of 1-forms,  
$d\hat x^{\hat\mu}=\hat e_{\nu}^{\hat\mu}dx^{\nu}$, we get the  
components of the metric tensor of the natural frame, 
\begin{equation}\label{gmunu}
g_{\mu \nu}=\eta_{\hat\alpha\hat\beta}\hat e^{\hat\alpha}_{\mu}\hat 
e^{\hat\beta}_{\nu}\,,\quad 
g^{\mu \nu}=\eta^{\hat\alpha\hat\beta} e_{\hat\alpha}^{\mu} 
e_{\hat\beta}^{\nu}\,. 
\end{equation} 
These raise or lower the {\em natural} vector indices, i.e., the Greek ones 
ranging from 0 to 3, while for the {\em local} vector indices, denoted by hat 
Greeks and having the same range, we must use the Minkowski metric. The 
derivatives  
$\hat\partial_{\hat\nu}=e^{\mu}_{\hat\nu}\partial_{\mu}$ satisfy the 
commutation rules  
\begin{equation}
[\hat\partial_{\hat\mu},\hat\partial_{\hat\nu}]
=e_{\hat\mu}^{\alpha} e_{\hat\nu}^{\beta}(\hat e^{\hat\sigma}_{\alpha,\beta}-
\hat e^{\hat\sigma}_{\beta,\alpha})\hat\partial_{\hat\sigma}
=C_{\hat\mu \hat\nu 
\cdot}^{~\cdot \cdot \hat\sigma}\hat\partial_{\hat\sigma}
\end{equation}
defining the Cartan coefficients which halp us to write the {\em conecttion} 
components in the local frames as   
\begin{equation}
\hat\Gamma^{\hat\sigma}_{\hat\mu \hat\nu}=e_{\hat\mu}^{\alpha}
e_{\hat\nu}^{\beta}(\hat e_{\gamma}^{\hat\sigma}
\Gamma^{\gamma}_{\alpha \beta}
-\hat e^{\hat\sigma}_{\beta, \alpha})=
\frac{1}{2}\eta^{\hat\sigma \hat\lambda}(C_{\hat\mu \hat\nu \hat\lambda}+
C_{\hat\lambda \hat\mu \hat\nu}+C_{\hat\lambda \hat\nu \hat\mu})\,.
\end{equation}
The notation  $\Gamma^{\gamma}_{\alpha \beta}$ stands for the usual 
Christoffel symbols involved in the formulas of the covariant derivatives 
$\nabla_{\mu}=~_{;\mu}$\,.

The Minkowski metric $\eta_{\hat\mu\hat\nu}$ remains invariant under the 
transformations of the {\em gauge} group of this metric, $G(\eta)=O(3,1)$. 
This has as subgroup the Lorentz group, $L_{+}^{\uparrow}$, of the 
transformations $\Lambda[A(\omega)]$ corresponding to the transformations 
$A(\omega)\in SL(2,C)$ through the canonical homomorphism \cite{W}. In the 
standard {\em covariant} parameterization, with the real parameters 
$\omega^{\hat\alpha \hat\beta}=-\omega^{\hat\beta\hat\alpha}$, we have
\begin{equation}\label{Aomega}
A(\omega)= e^{-\frac{i}{2}\omega^{\hat\alpha\hat\beta}
S_{\hat\alpha\hat\beta}}\,, 
\end{equation}
where $S_{\hat\alpha \hat\beta}$ are the covariant basis-generators of the 
$sl(2,C)$ Lie algebra which satisfy
\begin{equation}
\left[S_{\hat\mu\hat\nu},\,S_{\hat\sigma\hat\tau}\right]=i(
\eta_{\hat\mu\hat\tau}\,S_{\hat\nu\hat\sigma}-
\eta_{\hat\mu\hat\sigma}\,S_{\hat\nu\hat\tau}+
\eta_{\hat\nu\hat\sigma}\,S_{\hat\mu\hat\tau}-
\eta_{\hat\nu\hat\tau}\,S_{\hat\mu\hat\sigma})\,.
\end{equation}
For small values of $\omega^{\hat\alpha\hat\beta}$ the matrix elements of the 
transformations $\Lambda$ can be written as
\begin{equation}\label{infLam}
\Lambda[A(\omega)]^{\hat\mu\,\cdot}_{\cdot\,\hat\nu}=\delta^{\hat\mu}_{\hat\nu}
+\omega^{\hat\mu\,\cdot}_{\cdot\,\hat\nu}
+\cdots\,.
\end{equation}

Now we assume that $M$ is orientable and time-orientable such  
that $L^{\uparrow}_{+}$ can be considered as the gauge group of 
the Minkowski metric \cite{WALD}. In these conditions the  fields with spin 
can be defined as in the case of the flat spacetime, with the help of the 
finite-dimensional linear representations of the $SL(2,C)$ group 
\cite{W}. Then 
any field $\psi_{\rho}:M\to V_{\rho}$ is defined over $M$ with values in the 
vector space $V_{\rho}$ of the representation  $\rho$ which determines the 
spin content of $\psi_{\rho}$. In the following  we  systematically use the 
bases of $V_{\rho}$ labeled only by spinor or vector {\em local} indices 
defined with respect to the axes of the local frames given by the tetrad 
fields. Generally, these will not be written explicitly apart the cases when 
this is requested by the concrete calculation needs. 

The relativistic covariant field equations are derived from actions 
\cite{UK,SW},      
\begin{equation}\label{act}
{\cal S}[\psi_{\rho},e]=\int d^{4}x\sqrt{g}\,{\cal L}(
\psi_{\rho}, D_{\hat\mu}\psi_{\rho})\,,
\quad g=|\det(g_{\mu\nu})|\,, 
\end{equation}
with  covariant derivatives,  
\begin{equation}
D_{\hat\alpha}=\hat\partial_{\hat\alpha}+\frac{i}{2}\,
\rho(S^{\hat\beta\, \cdot}
_{\cdot\, \hat\gamma})\,\hat\Gamma^{\hat\gamma}_{\hat\alpha \hat\beta}\,,
\end{equation}
that assure  the invariance of the whole theory under the {\em gauge} 
transformations, 
\begin{eqnarray}
\hat e^{\hat\alpha}_{\mu}(x)&\to& \hat e'^{\hat\alpha}_{\mu}(x)=
\Lambda[A(x)]^{\hat\alpha\,\cdot}_{\cdot\,\hat\beta}
\,\hat e^{\hat\beta}_{\mu}(x)\,,\nonumber\\
e_{\hat\alpha}^{\mu}(x)&\to&  {e'}_{\hat\alpha}^{\mu}(x)=
\Lambda[A(x)]_{\hat\alpha\,\cdot}^{\cdot\,\hat\beta}
\,e_{\hat\beta}^{\mu}(x)\,,\label{gauge}\\
\psi_{\rho}(x)&\to&~\psi_{\rho}'(x)=\rho[A(x)]\,\psi_{\rho}(x)\,,\nonumber
\end{eqnarray}
determined by the mappings $A:M\to SL(2,C)$ the values of which are the 
local $SL(2,C)$ transformations $A(x)\equiv A[\omega(x)]$. These mappings can 
be organized as a group, ${\cal G}$, with respect to the multiplication  
$~\times ~$ defined as $(A'\times A)(x)=A'(x)A(x)$. The notation 
$Id$ stands for the mapping identity, $Id(x)=1\in SL(2,C)$, while $A^{-1}$ 
is the inverse of $A$, $(A^{-1})(x)=[A(x)]^{-1}$.

\subsection{Combined transformations}
\

The general coordinate transformations are automorphysms of $M$ which, in 
the passive mode, can be seen as changes of the local charts corresponding 
to the same domain of $M$ \cite{WALD,ON}. If $x$ and $x'$ are the coordinates 
of a  point in two different charts then there 
is a mapping $\phi$ between these charts giving the coordinate transformation, 
$x\to x'=\phi(x)$. These transformations  form a group with respect to the 
composition of mappings, $\,\circ\,$, defined as usual, i.e.  
$(\phi'\circ\phi)(x)=\phi'[\phi(x)]$. We denote this group by ${\cal A}$, 
its identity map by $id$ and the inverse mapping of $\phi$ by $\phi^{-1}$. 

The automothysms change all the components carrying  natural indices 
including those of the tetrad fields \cite{SW}. This means that, from the 
physical point of view, these transformations may change the positions of 
the local frames with respect to the natural ones. If we assume that the  
physical experiment makes reference to the axes of the local frame then it 
could 
appear situations when several correction of the positions of the local 
frames  should be needed before (or after) each general coordinate 
transformation. Obviously, these have to be done with the help of suitable 
gauge transformation  associated to the authomorphysms. Thus we arrive to the 
necessity of introducing  the {\em combined} transformations denoted by 
$(A,\phi)$ and defined as gauge transformations, given by $A\in {\cal G}$, 
followed by automorphysms, $\phi\in{\cal A}$. In this new notation the pure 
gauge transformations will appear as $(A,id)$ while the automorphysms will be 
denoted from now by $(Id,\phi)$.   

The  effect of a combined transformation $(A,\phi)$ upon our basic fields, 
$\psi_{\rho},\, e$ and $\hat e$ is 
$x\to x'=\phi(x),\,e(x)\to 
e'(x'),\, \hat e(x)\to \hat e'(x')$ and $\psi_{\rho}(x)\to \psi'_{\rho}(x')
=\rho[A(x)]\psi_{\rho}(x)$ where $e'$ are the transformed tetrads of the 
components
\begin{equation}
e'^{\mu}_{\hat\alpha}[\phi(x)]=\Lambda[A(x)]^{\cdot\,\hat\beta}
_{\hat\alpha\,\cdot}\,e^{\nu}_{\hat\beta}(x)
\frac{\partial\phi^{\mu}(x)}{\partial x^{\nu}}\,,
\end{equation} 
while the components of $\hat e'$ have to be calculated according to 
Eqs.(\ref{(duale)}). Thus we have written down the most general 
transformation laws that leave invariant the action in the sense that  
${\cal S}[\psi_{\rho}',e']={\cal S}[\psi_{\rho},e]$. The field equations 
derived from ${\cal S}$, written in operator form as 
$(E_{\rho}\psi_{\rho})(x)=0$, {\em covariantly} transform under these 
transformations,
\begin{equation}\label{ero}
(E_{\rho}\psi_{\rho})(x) \to
(E'_{\rho}\psi'_{\rho})(x')=\rho[A(x)](E_{\rho}\psi_{\rho})(x)\,,
\end{equation}
since the operators $E_{\rho}$ involve covariant derivatives \cite{SW}.

The association among the transformations of the groups ${\cal G}$ and 
${\cal A}$ must lead to a new group with a specific multiplication.  In order 
to find how looks this new operation  it is convenient to introduce the 
compositions among  the  mappings $A$ and $\phi$ (taken only in this order)   
giving new mappings, $A\circ\phi$, defined as $(A\circ \phi)(x)=A[\phi(x)]$. 
The  calculation rules  $Id\circ \phi=Id$, $A\circ id=A$ and 
$(A'\times A)\circ \phi=(A'\circ \phi)\times (A\circ \phi)$ are obvious.  
With these ingredients we define the new multiplication
\begin{equation}\label{comp}
(A',\phi')*(A,\phi)=
\left((A'\circ\phi)\times A,\phi'\circ\phi\right)\,.
\end{equation}
It is clear that now the identity is $(Id,id)$  while the inverse of a pair 
$(A,\phi)$ reads 
\begin{equation}\label{compAphi}
(A,\phi)^{-1}=(A^{-1}\circ\phi^{-1},\phi^{-1})\,.
\end{equation}    
First of all we observe that the operation $~*~$ is well-defined and 
represents 
the composition among the combined transformations since these can be  
expressed, according to their definition, as  $(A,\phi)=(Id,\phi)*(A,id)$.  
Furthermore,  we can convince ourselves that if we perform successively two 
arbitrary combined transformations, $(A,\phi)$ and $(A',\phi')$, then 
the resulting transformation is just $(A',\phi')*(A,\phi)$ as given by 
Eq.(\ref{comp}). This means that the combined transformations form a group 
with respect to the multiplication $\,*\,$.  
It is not difficult to verify that this group, denoted by $\tilde{\cal G}$, 
is the semidirect product $\tilde{\cal G}={\cal G}\circledS{\cal A}$ where 
${\cal G}$ is the {\em invariant} subgroup while ${\cal A}$ is an usual one. 

In the theories involving only vector and tensor fields we do not need to use   
the combined transformations  defined above since the theory is independent 
on the positions of the local frames. This can be easily shown even in our 
approach where we use field components with local indices. Indeed, if we 
perform a  combined transformation $(A,\phi)$ then  any tensor field  of 
rank  $(p,q)$,  
\begin{equation}
\psi^{\hat\alpha_{1}, \hat\alpha_{2},...,\hat\alpha_{p}} 
_{\hat\beta_{1}, \hat\beta_{2},...,\hat\beta_{q}}= 
\hat e^{\hat\alpha_{1}}_{\mu_{1}}\cdots
\hat e^{\hat\alpha_{p}}_{\mu_{p}}\,
 e_{\hat\beta_{1}}^{\nu_{1}}\cdots 
 e_{\hat\beta_{q}}^{\nu_{q}}\, 
\psi^{\mu_{1},\mu_{2},...,\mu_{p}}
_{\nu_{1},\nu_{2},...,\nu_{q}}\,,
\end{equation}
transforms according to the representation 
\begin{equation}
\rho_{\hat\alpha_{1}, \hat\alpha_{2},...,\hat\alpha_{p};\,  
\hat\beta'_{1}, \hat\beta'_{2},...,\hat\beta'_{q}} 
^{\hat\beta_{1}, \hat\beta_{2},...,\hat\beta_{q};\,
\hat\alpha'_{1}, \hat\alpha'_{2},...,\hat\alpha'_{p}}(A)= 
\Lambda^{\hat\beta_{1}\,\cdot}_{\cdot\,\hat\beta'_{1}}(A)\cdots\,  
\Lambda^{\cdot\, \hat\alpha'_{1}}_{\hat\alpha_{1}\,\cdot}(A)\cdots\,,  
\end{equation}
such that the resulting transformation law of the components carrying natural 
indices, 
\begin{equation}    
\psi'^{\mu_{1},...}
_{\,\nu_{1},...}(x')=
\frac{\partial x'^{\mu_{1}}}{\partial x^{\sigma_{1}}}\cdots\,
\frac{\partial x^{\tau_{1}}}{\partial x'^{\nu_{1}}}\cdots\,
\psi^{\sigma_{1},....}_{\tau_{1},....}(x)\,,
\end{equation}
is just the familiar one \cite{SW}. In other words, in this case the effect 
of the combined transformations reduces to that of their authomorohysms. 
However, when the half integer spin fields are  involved  this is no 
more true and we must use the combined transformations of  $\tilde{\cal G}$
if we want to keep under control the positions of the local frames.

\section{External symmetry}
\

In general, the symmetry of any manifold $M$ is given by its isometry group 
whose transformations leave invariant the metric tensor in any chart. The 
scalar field transforms under isometries according to the standard scalar 
representation generated by the orbital generators related to the Killing 
vectors of $M$ \cite{SW,WALD,ON}. In the following we propose a possible 
generalization of this theory of symmetry to the fields with spin, defining 
the external symmetry group and its representations.      

\subsection{Isometries}
\

There are conjectures when several  
coordinate transformations, $x\to x'=\phi_{\xi}(x)$, depend on $N$ 
independent real parameters, $\xi^a$ ($a,b,c...=1,2,...,N$),   
such that $\xi=0$ corresponds to the identity map, $\phi_{\xi=0}=id$. 
The set of these mappings  can be organized as a Lie group \cite{GIL}, 
$G\in {\cal G}$, if they accomplish the composition rule
\begin{equation}\label{compphi}
\phi_{\xi'}\circ \phi_{\xi}=\phi_{f(\xi',\xi)}\,,  
\end{equation}
where the functions $f: G\times G\to G$  define the group multiplication. 
These must satisfy $f^{a}(0,\xi)=f^{a}(\xi,0)=\xi^{a}$ and 
 $f^{a}(\xi^{-1},\xi)=f^{a}(\xi,\xi^{-1})=0$ where $\xi^{-1}$ are the 
parameters of the inverse mapping of $\phi_{\xi}$,  
$\phi_{\xi^{-1}}=\phi^{-1}_{\xi}$. 
Moreover, the structure constants of $G$ can be calculated as \cite{HAM}
\begin{equation}\label{c}
c_{abc}=\left(\frac{\partial f^{c}(\xi,\xi')}{\partial \xi^{a}\partial \xi'^{b}}-
\frac{\partial f^{c}(\xi,\xi')}{\partial \xi^{b}\partial \xi'^{a}}
\right)_{|\xi=\xi'=0}\,.
\end{equation}
For small values of the group parameters the infinitesimal transformations,  
$x^{\mu}\to x'^{\mu}=x^{\mu}+\xi^{a}k_{a}^{\mu}(x)+\cdots$, 
are given by the vectors $k_{a}$ whose components,
\begin{equation}\label{ka}
k_{a}^{\mu}={\frac{\partial \phi^{\mu}_{\xi}}{\partial\xi^{a}}}_{|\xi=0}\,,
\end{equation}
satisfy the  identities
\begin{equation}\label{kkc}
k^{\mu}_{a}k^{\nu}_{b,\mu}
-k^{\mu}_{b}k^{\nu}_{a,\mu}+c_{abc}k^{\nu}_{c}=0\,,
\end{equation}
resulting from Eqs.(\ref{compphi}) and (\ref{c}).

In the following we restrict ourselves to consider only the {\em isometry} 
transformations, $x'=\phi_{\xi}(x)$, which leave invariant the components of 
the metric tensor \cite{SW,ON}, i.e.   
\begin{equation}\label{giso}
g_{\alpha\beta}(x')\frac{\partial x'^\alpha}{\partial x^\mu}
\frac{\partial x'^\beta}{\partial x^\nu}=g_{\mu\nu}(x)\,.
\end{equation}
These form the isometry group $G\equiv I(M)$ which is the Lie group giving the 
symmetry of the spacetime $M$. We consider that this has $N$ independent 
parameters and, therefore,  $k_{a}, a=1,2,...N$, are independent Killing 
vectors (which satisfy $k_{a\, \mu;\nu}+k_{a\, \nu;\mu}=0$). Then their 
corresponding Lie derivatives form a basis of the Lie algebra $i(M)$ of the 
group $I(M)$ \cite{ON}.  

However, in practice we are interested to find the operators of the 
relativistic quantum 
theory related to these geometric objects which describe the symmetry of 
the background. For this reason we focus upon the operator-valued 
representations \cite{BR} of the group $I(M)$ and its 
algebra. The {\em scalar} field $\psi:M\to C$ transforms under isometries as
$\psi(x)\to \psi'[\phi_{\xi}(x)]=\psi(x)$. This rule defines the 
representation $\phi_{\xi}\to T_{\xi}$ of the group $I(M)$ whose 
operators have the action $\psi'=T_{\xi}\psi=\psi\circ\phi^{-1}_{\xi}$. 
Hereby it results that the operators of  
infinitesimal transformations, $T_{\xi}=1-i\xi^{a}L_{a}+\cdots$, depend on 
the basis-generators,     
\begin{equation}\label{genL}
L_{a}=-ik_{a}^{\mu}\partial_{\mu}\,, \quad a=1,2,...,N\,,
\end{equation}
which are completely determined by the Killing vectors. From Eq.(\ref{kkc}) we 
see that they obey the commutation rules 
\begin{equation}\label{comL}
[L_{a}, L_{b}]=ic_{abc}L_{c}\,,
\end{equation} 
given by the structure constants of $I(M)$. In other words they form a basis 
of the operator-valued representation of the Lie algebra $i(M)$ in a carrier 
space of scalar fields. Notice that in the usual quantum mechanics the 
operators similar to the generators $L_{a}$ are called often {\em orbital} 
generators.  

\subsection{The group of external symmetry}
\

Now  the problem is  how may transform  under isometries the whole geometric 
framework of the theories with spin where we explicitly use the local frames.
Since the isometry is a general coordinate transformation it changes the 
relative positions of the local and natural frames. This fact may be an 
impediment when one intends to study the symmetries of the theories with spin 
induced by  those of the background. For this reason it is natural to 
suppose that the good symmetry  transformations we need are combined 
transformations in which the isometries are preceded by appropriate gauge 
transformations such that not only  the form of the metric tensor should be 
conserved but the form of the tetrad field components too.

Thus we arrive at the main point of our proposal. We introduce the {\em external 
symmetry} transformations, $(A_{\xi},\phi_{\xi})$, as combined transformations 
involving isometries and  corresponding gauge transformations necessary to 
{\em preserve the gauge}. We assume that in a fixed gauge, given by the 
tetrad fields $e$ and $\hat e$, $A_{\xi}$ is defined by
\begin{equation}\label{Axx}
\Lambda[A_{\xi}(x)]^{\hat\alpha\,\cdot}_{\cdot\,\hat\beta}=
\hat e_{\mu}^{\hat\alpha}[\phi_{\xi}(x)]\frac{\partial \phi^{\mu}_{\xi}(x)}
{\partial x^{\nu}}\,e^{\nu}_{\hat\beta}(x)\,, 
\end{equation}    
with the supplementary condition $A_{\xi=0}(x)=1\in SL(2,C)$. 
Since $\phi_{\xi}$ is an isometry Eq.(\ref{giso}) guarantees that 
$\Lambda[A_{\xi}(x)]\in L^{\uparrow}_{+}$ and, implicitly, 
$A_{\xi}(x)\in SL(2,C)$. Then the transformation laws of our fields are
\begin{equation}\label{es}
(A_{\xi},\phi_{\xi}):\qquad
\begin{array}{rlrcl}
x&\to&x'&=&\phi_{\xi}(x)\,,\\
e(x)&\to&e'(x')&=&e[\phi_{\xi}(x)]\,,\\
\hat e(x)&\to&\hat e'(x')&=&\hat e[\phi_{\xi}(x)]\,,\\
\psi_{\rho}(x)&\to&\psi_{\rho}'(x')&=&\rho[A_{\xi}(x)]\psi_{\rho}(x)\,.
\end{array}
\qquad
\end{equation}
The mean virtue of these transformations are that they leave {\em invariant} 
the operators of the field equations, $E_{\rho}$, since the components of 
the tetrad fields and, consequently, the covariant derivatives do not change 
their form.

For small $\xi^{a}$ the covariant $SL(2,C)$  parameters 
of $A_{\xi}(x)\equiv A[\omega_{\xi}(x)]$ can be written as 
$\omega^{\hat\alpha\hat\beta}_{\xi}(x)= 
\xi^{a}\Omega^{\hat\alpha\hat\beta}_{a}(x)+\cdots$ where, 
according to  Eqs.(\ref{Aomega}), (\ref{infLam}) 
and (\ref{Axx}), we have
\begin{equation}\label{Om}     
\Omega^{\hat\alpha\hat\beta}_{a}\equiv {\frac{\partial 
\omega^{\hat\alpha\hat\beta}_{\xi}}
{\partial\xi^a}}_{|\xi=0}
=\left( \hat e^{\hat\alpha}_{\mu}\,k_{a,\nu}^{\mu}
+\hat e^{\hat\alpha}_{\nu,\mu}
k_{a}^{\mu}\right)e^{\nu}_{\hat\lambda}\eta^{\hat\lambda\hat\beta}\,.
\end{equation}  
We must specify that these functions are antisymmetric if and only if $k_{a}$ 
are Killing vectors. This indicates that the association among isometries 
and the gauge transformations defined by Eq.(\ref{Axx}) is correct.

It remains to show that the transformations 
$(A_{\xi},\phi_{\xi})$ form a Lie group related to 
$I(M)$. Starting with Eq.(\ref{Axx}) we find that 
\begin{equation}\label{compA}
(A_{\xi'}\circ\phi_{\xi})\times A_{\xi}=A_{f(\xi',\xi)}\,,
\end{equation}
and, according to Eqs.(\ref{compphi}) and (\ref{compA}), we obtain
\begin{equation}\label{mult}
(A_{\xi'},\phi_{\xi'})*(A_{\xi},\phi_{\xi})=
(A_{f(\xi',\xi)},\phi_{f(\xi',\xi)})\,,
\end{equation}  
and $(A_{\xi=0},\phi_{\xi=0})=(Id,id)$.
Thus we have shown that the  pairs $(A_{\xi},\phi_{\xi})$ 
form a Lie group with respect to the operation $~*~$. 
We say that this is the  external symmetry group of $M$ and we denote it 
by $S(M)\subset \tilde{\cal G}$. From Eq.(\ref{mult}) we understand that $S(M)$ 
is {\em locally 
isomorphic} with $I(M)$ and, therefore, the Lie algebra of $S(M)$, denoted by 
$s(M)$, is isomorphic with $i(M)$ having the same structure constants.  
In our opinion, $S(M)$ must be isomorphic with the universal covering group of 
$I(M)$ since it has  anyway the topology induced by $SL(2,C)$ which is simply 
connected. 
In general, the number of group parameters of $I(M)$ or $S(M)$ 
(which is equal to the number of the independent Killing vectors of $M$) 
can be $0\le N\le 10$.

The form of the external symmetry transformations is strongly dependent on the 
choice of the local chart as well as that of the tetrad gauge. If 
we change simultaneously the gauge and the coordinates with the help of a 
combined transformation $(A,\phi)$ then each 
$(A_{\xi},\phi_{\xi})\in S(M)$ transforms as
\begin{equation}\label{aaprim}
(A_{\xi},\phi_{\xi})\to 
(A'_{\xi},\phi'_{\xi})=(A,\phi)*(A_{\xi},\phi_{\xi})*(A,\phi)^{-1} 
\end{equation}  
which means that
\begin{eqnarray}\label{apsiprim}
A'_{\xi}&=&\left\{\left[\left(A\circ\phi_{\xi}\right)\times A_{\xi}\right]
\times A^{-1}\right\}\circ \phi^{-1}\,,\\
\phi'_{\xi}&=&\left(\phi\circ\phi_{\xi}\right)\circ\phi^{-1}\,.
\end{eqnarray}

\subsection{Representations}
\

The last of Eqs.(\ref{es}) which gives the transformation law of the field 
$\psi_{\rho}$ defines the operator-valued representation  
$(A_{\xi},\phi_{\xi})\to T_{\xi}^{\rho}$ of the group $S(M)$,   
\begin{equation}\label{rep}
(T_{\xi}^{\rho}\psi_{\rho})[\phi_{\xi}(x)]=\rho[A_{\xi}(x)]\psi_{\rho}(x)\,.
\end{equation}
The invariance of the field equations under 
these transformations requires to have  
\begin{equation}\label{invE}
T^{\rho}_{\xi}E_{\rho}{(T^{\rho}_{\xi})}^{-1}=E_{\rho}\,.
\end{equation}
Since  $A_{\xi}(x)\in SL(2,C)$  we  say that this representation is 
{\em induced} by the representation $\rho$ of $SL(2,C)$ \cite{BR,MAK}. 
As we have shown in Sec.2.2, if $\rho$ is a vector or tensor representation 
(having only integer spin components) then the effect of the transformation    
(\ref{rep}) upon the components carrying natural indices is due only to 
$\phi_{\xi}$. However, for the representations with half integer spin the 
presence of $A_{\xi}$ is crucial since there are no natural indices. 
Moreover, this allows us to define  the generators of the 
representations (\ref{rep}) for any spin.  

The basis-generators of the representations of the Lie algebra 
$s(M)$ are the operators
\begin{equation}\label{X}
X^{\rho}_{a} = i{\frac{\partial T_{\xi}^{\rho}}{\partial \xi^{a}}}_{|\xi=0}=
S_{a}^{\rho}+L_{a}\,,
\end{equation}
which appear as  sums among the orbital generators defined by Eq.(\ref{genL}) 
and the {\em spin terms}  which have the action 
\begin{equation}\label{sss}
(S_{a}^{\rho}\psi_{\rho})(x)=\rho[S_{a}(x)]\psi_{\rho}(x)\,. 
\end{equation}
This is determined by the form of the {\em local} $sl(2,C)$ generators, 
\begin{equation}\label{Sx}
S_{a}(x)=i{\frac{\partial A_{\xi}(x)}{\partial \xi^{a}}}_{|\xi=0}=
\frac{1}{2}\Omega^{\hat\alpha\hat\beta}_{a}(x)
S_{\hat\alpha\hat\beta}\,,
\end{equation}
which depend on the functions (\ref{Om}).
Furthermore,  if we derive  Eq.(\ref{compA}) with respect to $\xi$ and $\xi'$ 
then from Eqs.(\ref{infLam}), (\ref{c}) and (\ref{Om}), after a few 
manipulations, we find the identities
\begin{equation}\label{idOM}
\eta_{\hat\alpha\hat\beta}\left(
\Omega_{a}^{\hat\alpha\hat\mu}\Omega_{b}^{\hat\beta\hat\nu}  
-\Omega_{b}^{\hat\alpha\hat\mu}\Omega_{a}^{\hat\beta\hat\nu}\right)+
k^{\mu}_{a}\Omega_{b,\mu}^{\hat\mu\hat\nu}
-k^{\mu}_{b}\Omega_{a,\mu}^{\hat\mu\hat\nu}+c_{abc}\Omega_{c}^{\hat\mu\hat\nu}
=0\,.
\end{equation}
Hereby it results that  
\begin{equation}
[S_{a}^{\rho},S_{b}^{\rho}]+[L_{a},S_{b}^{\rho}]-[L_{b},S_{a}^{\rho}]
=ic_{abc}S_{c}^{\rho}\,,
\end{equation}
and, according to Eq.(\ref{comL}), we find the expected commutation rules 
\begin{equation}\label{comX}
[X_{a}^{\rho}, X_{b}^{\rho}]=ic_{abc}X_{c}^{\rho}\,.
\end{equation} 
Thus we have obtained the basis-generators of a operator-valued representation  
of $s(M)$  induced by the linear representation $\rho$ of $sl(2,C)$. All 
the operators of this representation commute with the operator $E_{\rho}$ 
since, according to Eqs.(\ref{invE}) and (\ref{X}), we have 
\begin{equation}
[E_{\rho},X^{\rho}_{a}]=0\,, \quad a=1,2,...,N\,.
\end{equation}
Therefore, for defining quantum modes we can use the  set of commuting 
operators containing   the Casimir operators those of the Cartan subalgebra 
and $E_{\rho}$.

The action of the operators $X_{a}^{\rho}$ depends on the choice of many 
elements: the 
natural coordinates, the tetrad gauge, the group parameterization and the 
representation $\rho$. What is important here is that they are strongly  
dependent  on the tetrad gauge fixing even in the case of the representations 
with integer spin. This is because the covariant parameterization of the  
$sl(2,C)$ algebra is defined with respect to the axes of the 
local frames.  In general, if we consider the 
representation $(A_{\xi},\phi_{\xi})\to T^{\rho}_{\xi}$ 
and we perform a combined transformation (\ref{aaprim}) then it results the 
{\em equivalent} representation, $(A'_{\xi},\phi'_{\xi})\to T'^{\rho}_{\xi}$.  
Its generators  calculated  from Eqs.(\ref{apsiprim}) indicate that in this 
case the equivalence relations are much more complicated than those of the 
usual theory of linear representations.  
Without to enter in other technical details we specify that if we change only 
the gauge with the help of the transformation $(A, id)$ then the local 
$sl(2,C)$ generators (\ref{Sx}) transform as 
\begin{eqnarray}
S_{a}(x)\to S'_{a}(x)&=& A(x)S_{a}(x) A(x)^{-1}\nonumber\\
&&+k_{a}^{\sigma}(x)\Lambda[A(x)]_{\hat\alpha
\hat\mu,\,\sigma}\Lambda[A(x)]_{\hat\beta\,\cdot}^{\cdot\,\hat\mu}
S^{\hat\alpha\hat\beta}\,,
\end{eqnarray}
while the orbital parts do not change their form. This means that the gauge 
transformations change, in addition, the commutation relations among the spin 
and orbital parts of the generators $X_{a}^{\rho}$. Hence we can draw the 
conclusion that the choice of the tetrad gauge which defines the local frames 
may have important consequences upon the measurement of the local spin 
effects.  

There are gauge fixings where  the local $sl(2,C)$ generators 
$S_{a}(x)$, $a=1,2,...,n$ ($n\le N$), corresponding to a subgroup 
$H\subset S(M)$, are independent on $x$ and, therefore, 
$[S_{a}^{\rho},L_{b}]=0$ for all $a=1,2,...,n$  and $b=1,2,...,N$. Then the 
operators $S_{a}^{\rho}$  are just the basis-generators of an usual 
linear representation of $H$  and the field $\psi_{\rho}$ behaves 
{\em manifestly covariant} under the external symmetry transformations of this 
subgroup. Of course, when $H=S(M)$  we say simply that the field $\psi_{\rho}$ 
is manifest covariant. 

The simplest examples are the manifest covariant fields of special relativity. 
Here  since the spacetime $M$ is flat the metric in Cartesian coordinates is      
$g_{\mu\nu}=\eta_{\mu\nu}$ and one can use the {\em inertial} (local) frames
with $e^{\mu}_{\nu}=\hat e^{\mu}_{\nu}=\delta^{\mu}_{\nu}$. Then   
the isometries are just the transformations $x'=\Lambda[A(\omega)]x -a$ of 
the Poincar\' e group,
${\cal P}_{+}^{\uparrow} = T(4)\circledS L_{+}^{\uparrow}$  \cite{W}. 
If we denote by $\xi^{(\mu\nu)}=\omega^{\mu\nu}$ the $SL(2,C)$ parameters and 
by $\xi^{(\mu)}=a^{\mu}$ those of the translation group $T(4)$, then we 
find that  $S(M)\equiv \tilde{\cal P}^{\uparrow}_{+}=T(4)\circledS SL(2,C)$  
is just the universal covering group 
of ${\cal P}_{+}^{\uparrow}$. 
Furthermore, it is a simple exercise to calculate 
the basis-generators 
\begin{eqnarray}
X_{(\mu)}^{\rho}&=&i\partial_{\mu}\,,\\
X_{(\mu\nu)}^{\rho}&=&\rho(S_{\mu\nu})+i(\eta_{\mu\alpha}x^{\alpha}
\partial_{\nu}- \eta_{\nu\alpha}x^{\alpha}\partial_{\mu})\,, 
\end{eqnarray}
which show us that $\psi_{\rho}$ transforms manifestly covariant.

In general, there are many cases of curved spacetimes for which one can choose 
suitable local frames allowing one to introduce manifest covariant fields with 
respect to a subgroup $H\subset S(M)$ or even  the whole group $S(M)$. 
In our opinion, this is possible only when $H$ or $S(M)$ are at most subgroups 
of $\tilde{\cal P}_{+}^{\uparrow}$.

\section{The central symmetry}
\

Let us take as first example the spacetimes $M$ which have spherically 
symmetric static chart that will be referred here as {\em central} charts.   
These manifolds have the  isometry group $I(M)=T(1)\otimes SO(3)$ 
of time translations and space rotations.

\subsection{Central charts}
\

In a central chart with 
Cartesian coordinates $x^{0}=t$ and  $x^i$ ($i,j,k...=1,2,3$),  the metric 
tensor is time-independent and transforms manifestly covariant  under the  
rotations $R\in SO(3)$ of the space coordinates,
\begin{equation}\label{(rot)}  
t'=t,\quad  x'^{i}= R_{\cdot\,j}^{i\,\cdot}(\omega)x^{j}=x^{i}+
\omega^{i\,\cdot}_{\cdot\,j}x^{j}+\cdots\,,
\end{equation}
denoted simply by $x\to x'=Rx$. Here the most general form of the 
line element,       
\begin{equation}\label{(metr)} 
ds^{2}=g_{\mu\nu}(x)dx^{\mu}dx^{\nu}=A(r)dt^{2}-[B(r)\delta_{ij}+
C(r)x^{i}x^{j}]dx^{i}dx^{j}\,,
\end{equation} 
may involve three functions, $A$, $B$ and $C$, depending only on  the 
Euclidean norm of $\stackrel{\rightarrow}{x}$,  
$r=\vert\stackrel{\rightarrow}{x}\vert$. In applications it is convenient to 
replace these 
functions by  new ones,  $u$,  $v$ and $w$, defined as
\begin{equation}\label{(ABC)}   
A=w^{2}, \quad B=\frac{w^2}{v^2}, \quad 
C=\frac{w^2}{r^2}\left( \frac{1}{u^2}-\frac{1}{v^2}\right)\,.
\end{equation}

Other useful central charts are those with  spherical coordinates, 
$r$, $\theta$, $\phi$, commonly associated with the  Cartesian space ones. 
Here the line elements are   
\begin{equation}\label{(muvw)}
ds^{2}=w^{2}dt^{2}-\frac{w^2}{u^2}dr^{2}-
\frac{w^2}{v^2}r^{2}(d\theta^{2}+\sin^{2}\theta d\phi^{2})\,.
\end{equation}
In these charts we see that the advantage of the new functions we have 
introduced is of simple transformation laws under the isotropic dilatations 
which change only the radial coordinate, $r\to r'(r)$, without to affect the 
central symmetry of the line element. These transformations, 
\begin{equation}
u'(r')=u(r)\left|\frac{dr'(r)}{dr}\right|\,,\quad
v'(r')=v(r)\frac{r'(r)}{r}\,,\quad
w'(r')=w(r)\,, 
\end{equation}   
allow one to choose desired forms for the functions $u,v$ and $w$.

\subsection{The Cartesian gauge}
\

The Cartesian gauge in central charts was  mentioned  long time ago \cite{BW} 
but it is less used in concrete problems since it leads to complicated 
calculations in spherical coordinates. However, in Cartesian coordinates this 
gauge has the advantage of explicitly pointing out the global central symmetry 
of the manifold. In Refs.\cite{COTA} we have proposed a version of Cartesian gauge in 
central charts with Cartesian coordinates that preserve the manifest covariance under rotations 
(\ref{(rot)}) in the sense that the corresponding 1-forms transform as 
\begin{equation}\label{(tr)}
d\hat x ^{\hat\mu}\to d\hat x'^{\hat\mu}=\hat e^{\hat\mu}_{\alpha}(x')
dx'^{\alpha}=(Rd\hat x)^{\hat\mu}.
\end{equation}
If the line element has the form (\ref{(metr)}) then the most general choice 
of the tetrad fields with the above property is
\begin{eqnarray}
&&\hat e^{0}_{0}=\hat a(r), \quad \hat e^{0}_{i}=\hat e^{i}_{0}=0, \quad
\hat e^{i}_{j}=\hat b(r)\delta_{ij}+\hat c(r) x^{i}x^{j}
+\hat d(r) \epsilon_{ijk}x^{k},\label{(eee)}\\
&&e^{0}_{0}= a(r), \quad  e^{0}_{i}= e^{i}_{0}=0, \quad
e^{i}_{j}= b(r)\delta_{ij}+ c(r) x^{i}x^{j}
+ d(r) \epsilon_{ijk}x^{k}\,,\label{(eee1)}
\end{eqnarray}
where, according to (\ref{gmunu}), (\ref{(metr)}) and (\ref{(ABC)}), we must 
have 
\begin{eqnarray}
&&\hat a=w\,,\, ~ \hat b=\frac{w}{v}\cos\alpha\,,~\hat c=\frac{1}{r^2}
\left( \frac{w}{u}-\frac{w}{v}\cos\alpha\right),~
\hat d=\frac{1}{r}\frac{w}{v}\sin\alpha\,, \label{(abc)}\\
&&a= \frac{1}{w}\,, ~  b=\frac{v}{w}\cos\alpha\,, ~c=\frac{1}{r^2}
\left( \frac{u}{w}-\frac{v}{w}\cos\alpha\right),~
d=-\frac{1}{r}\frac{v}{w}\sin\alpha\,. 
\label{(abc1)}
\end{eqnarray}

The angle  $\alpha$ is an arbitrary function of $r$ which is not explicitly 
involved in the expression of the metric tensor since it represents the angle 
of an arbitrary rotation of the local frame around the direction  of 
$\stackrel{\rightarrow}x$, that does not change the relative position 
of $\stackrel{\rightarrow}x$ with respect to this frame. 

When one defines the metric tensor such that 
{${g_{\mu\nu}}_{|r=0}=\eta_{\mu\nu}$ then  $u(0)^{2}=v(0)^{2}=w(0)^{2}=1$. 
Moreover, it is natural to take $\alpha(0)=0$. In other respects, from 
Eqs.(\ref{(abc)}) and (\ref{(abc1)}) we 
see that the function $w$ must be positively defined in order to keep the 
same sense for the time axes of the natural and local frames. In addition, it 
is convenient to consider that the function $u$ is positively defined too. 
However, the function $v=\eta_{P}|v|$ has the sign given by the relative 
parity $\eta_{P}$ which takes the value  $\eta_{P}=1$ when the space axes of 
the local frame at $x=0$ are parallel with those of the natural frame, and  
$\eta_{P}=-1$ if these are antiparalel.

Now we have all the elements we need to calculate the generators of the 
representations $T^{\rho}$ of the group $S(M)$. If we denote by $\xi^{(0)}$ 
the parameter of the 
time translations and by $\xi^{(i)}=\epsilon_{ijk}\omega^{jk}/2$ the parameters 
of the rotations (\ref{(rot)}), we find  that the local $sl(2,C)$ generators 
of Eq.(\ref{Sx}) are just the $su(2)$ ones, i.e. $S_{(ij)}(x)=S_{ij}$, such 
that the basis-generators read 
\begin{equation}\label{ang1}
X_{(0)}^{\rho}=i\partial_{t}\,,\quad
X_{(i)}^{\rho}=
\frac{1}{2}\epsilon_{ijk}\,\rho(S_{jk}) 
+L_{(i)}
\end{equation}
where $L_{(i)}=-i\epsilon_{ijk}x^{j}\partial_{k}$ are the usual components of 
the orbital angular momentum. Thus we obtain that the group $S(M)=T(1)\otimes   
SU(2)$ is the universal covering group of $I(M)$. The physical 
significance of the basis-generators is the usual one, namely  
$X_{(0)}^{\rho}$ is the Hamiltonian operator while  
$X_{(i)}^{\rho}\equiv J^{\rho}_{(i)}$ are the components of the whole 
angular momentum operator of the field 
$\psi_{\rho}$  which transforms manifestly covariant. 

We can conclude that, in our Cartesian gauge, the local frames play the same 
role as the usual Cartesian {\rm rest frames} of the central sources in flat 
spacetime since their axes are just those of projections of the angular 
momenta. 

\subsection{The diagonal gauge}
\

In other gauge fixings the basis-generators are quite 
different. A tetrad gauge largely used in central charts with spherical 
coordinates \cite{D} is the diagonal  gauge defined  by the 1-forms
\begin{equation}\label{1fsf}
d\hat x^{0}_{s}=wdt\,,\quad
d\hat x^{1}_{s}=\frac{w}{u}dr\,,\quad
d\hat x^{2}_{s}=r\frac{w}{v}d\theta\,,\quad
d\hat x^{3}_{s}=r\frac{w}{v}\sin\theta d\phi\,.
\end{equation}
In this gauge the angular momentum operators of the canonical basis 
(where $J_{(\pm)}=J_{(1)}\pm iJ_{(2)}$) are
\begin{equation}
J^{\rho}_{(\pm)}=\frac{e^{\pm i\phi}}{\sin\theta}\,\rho(S_{23})+L_{(\pm)}\,,\quad 
J^{\rho}_{(3)}=L_{(3)}\,.
\end{equation}
Thus we obtain a representation of $SU(2)$ where the spin terms do not 
commute with the orbital ones and, therefore, the field $\psi_{\rho}$ does not 
transform manifestly covariant under rotations. In this case we can say that 
the spin part of the central symmetry remains partially hidden  because 
of the diagonal gauge which determines  special positions of the local frames 
with respect to the natural one. However, when this is an impediment one can 
change anytime this gauge into the Cartesian one by using a simple local 
rotation. For the flat spacetimes these transformations and their effects 
upon the Dirac equation are studied in Ref.\cite{VIL}. We note that the form 
of the spin generators as well as that of the mentioned rotation depend on 
the enumeration of the 1-forms (\ref{1fsf}).   

\newpage

\section{The dS and AdS symmetries}
\

The backgrounds with highest external symmetry are the dS and the AdS 
spacetimes. We shall briefly discuss simultaneously both these manifolds 
which will be denoted by $M_{\epsilon}$ where $\epsilon=1$ for dS case 
and $\epsilon=-1$ for AdS one. Our goal here is to calculate the  generators  
of the representations of the group $S(M_{\epsilon})$ 
induced by those of $SL(2,C)$.

The dS and AdS spacetimes are hyperboloids in the $(4+1)$ or 
$(3+2)$-dimensional flat spacetimes, $M_{\epsilon}^{5}$, of coordinates 
$Z^{A},\, A,B,...=0,1,2,3,5$, and the metric 
$\eta(\epsilon)={\rm diag}(1,-1,-1,-1,-\epsilon)$. The equation of the 
hyperboloid of radius $r_{0}=1/\hat\omega$  reads
\begin{equation}\label{hip}
-\eta_{AB}(\epsilon)Z^{A}Z^{B}=\epsilon\,{r_{0}}^{2}\,.
\end{equation}
From their definitions it results that the dS or AdS spacetimes are 
homogeneous spaces of the pseudo-orthogonal groups $SO(4,1)$ or $SO(3,2)$ 
which play the role of gauge groups of the metric $\eta(\epsilon)$ (for 
$\epsilon=1$ and  $\epsilon =-1$ respectively) and represent just the 
isometry groups of these manifolds, $G[\eta(\epsilon)]=I(M_{\epsilon})$. 
Then it is natural to use the  {\em covariant}  real parameters 
$\omega^{AB}=-\omega^{BA}$ since in this parameterization the orbital 
basis-generators of the representations of $G[\eta(\epsilon)]$ carried by the 
spaces of the  functions over $M_{\epsilon}^{5}$ have the   
usual form
\begin{equation}\label{LAB5}
 L_{AB}^{5}=i\left[\eta_{AC}(\epsilon)Z^{C}\partial_{B}-
 \eta_{BC}(\epsilon)Z^{C}
\partial_{A}\right].
\end{equation} 
They  will give us directly the orbital basis-generators of the  
representations of $S(M_{\epsilon})$ in the carrier spaces of the 
functions defined over dS or AdS spacetimes.

\subsection{Central charts}
\ 

The hyperboloid equation can be solved  
in Cartesian dS/AdS coordinates, $x^{0}=t$ and $x^{i}$ ($i=1,2,3$), which 
satisfy    
\begin{eqnarray}\label{dScart}
Z^{5}\!\!&=&\hat\omega^{-1}\chi_{\epsilon}(r)\left\{\begin{array}{lll}
\cosh \hat\omega t&{\rm if}&\epsilon=1\\
\cos \hat\omega t&{\rm if}&\epsilon=-1
\end{array}\right.\nonumber\\
Z^{0}\!\!&=&\hat\omega^{-1}\chi_{\epsilon}(r)\left\{\begin{array}{lll}
\sinh \hat\omega t&{\rm if}&\epsilon=1\\
\sin \hat\omega t&{\rm if}&\epsilon=-1
\end{array}\right.\label{dsadsx}\\
Z^{i}&=& x^{i}\,,\nonumber 
\end{eqnarray} 
where we have denoted 
$\chi_{\epsilon}(r)=\sqrt{1-\epsilon\,\omega^{2}r^2}$.
The line elements  
\begin{eqnarray}\label{dSmet}
ds^{2}&=&\eta_{AB}(\epsilon)dZ^{A}dZ^{B}\label{(adsm)}\\
&=&\chi_{\epsilon}(r)^{2} dt^{2}-
\frac{dr^{2}}{\chi_{\epsilon}(r)^{2}} -
 r^{2}(d\theta^{2}+\sin^{2}\theta\,d\phi^{2})\,,\nonumber
\end{eqnarray}
are defined on  the radial domains $D_{r}=[0,1/\sqrt{\hat\omega})$ or 
$D_{r}=[0,\,\infty)$ for dS or AdS respectively. 

We calculate the Killing vectors and the orbital generators of the external 
symmetry in the Cartesian coordinates defined by Eq.(\ref{dScart}) and 
the mentioned  parameterization of $I(M_{\epsilon})$ starting with the 
identification $\xi^{(AB)}=\omega^{AB}$. Then, from Eqs.(\ref{genL}) and  
(\ref{LAB5}), after a few manipulations, we obtain the orbital 
basis-generators 
\begin{eqnarray}
L_{(05)}&=&\frac{i\epsilon}{\hat\omega}\,\partial_{t}\,,\\
L_{(j5)}&=&\frac{i\epsilon}{\hat\omega}\chi_{\epsilon}(r)\,
\left( \begin{array}{l}
{\rm cosh}\,\hat\omega t\\
\cos\hat\omega t
\end{array}\right)
\partial_{j}+\frac{ix^{j}}{\chi_{\epsilon}(r)}\, 
\left(\begin{array}{l}
{\rm sinh}\,\hat\omega t\\
\sin\hat\omega t
\end{array}\right)
\partial_{t}\,,\\ 
L_{(0j)}&=&\frac{i}{\hat\omega}\chi_{\epsilon}(r)
\left(\begin{array}{l}
{\rm sinh}\,\hat\omega t\\
\sin\hat\omega t
\end{array}\right)
\partial_{j}+\frac{ix^{j}}{\chi_{\epsilon}(r)}
\left(\begin{array}{l}
{\rm cosh}\,\hat\omega t\\
\cos\hat\omega t 
\end{array}\right)
\partial_{t}\,,\\ 
L_{(ij)}&=&-i\,(x^{i}\,\partial_{j}-x^{j}\,\partial_{i})\,.
\end{eqnarray}
Furthermore, we consider the Cartesian tetrad gauge defined by 
Eqs.(\ref{(eee)})  - (\ref{(abc1)}) where, according to Eq.(\ref{dSmet}), 
we have    
\begin{equation}\label{dSuvw}
u(r)=\chi_{\epsilon}(r)^{2}\,,\quad 
v(r)=w(r)=\chi_{\epsilon}(r)\,.
\end{equation}
In addition  we take $\alpha=0$. In this gauge we obtain the following local 
$sl(2,C)$ generators   
\begin{eqnarray}
S_{(05)}(x)&=&0\,,\\
S_{(j5)}(x)&=&S_{0j}
\left(\begin{array}{l}
{\rm sinh}\,\hat\omega t\\
\sin\hat\omega t 
\end{array}\right)
+\frac{1}{r^2}[\chi_{\epsilon}(r)
-1]\left[\epsilon\frac{S_{jk}x^{k}}{\hat\omega}
\left(\begin{array}{l}
{\rm cosh}\,\hat\omega t\\
\cos\hat\omega t
\end{array}\right)
\right.\nonumber\\
&&\left.
-\frac{S_{0k}x^{k}x^{j}}{\chi_{\epsilon}(r)}
\left(\begin{array}{l}
{\rm sinh}\,\hat\omega t\\
\sin\hat\omega t 
\end{array}\right)
\right]\,,\\ 
S_{(0j)}(x)&=&S_{0j}
\left(\begin{array}{l}
{\rm cosh}\,\hat\omega t\\
\cos\hat\omega t 
\end{array}\right)
+\frac{1}{r^2}[\chi_{\epsilon}(r)
-1]\left[\frac{S_{jk}x^{k}}{\hat\omega}
\left(\begin{array}{l}
{\rm sinh}\,\hat\omega t\\
\sin\hat\omega t
\end{array}\right)
\right.\nonumber\\
&&\left.
-\frac{S_{0k}x^{k}x^{j}}{\chi_{\epsilon}(r)}
\left(\begin{array}{l}
{\rm cosh}\,\hat\omega t\\
\cos\hat\omega t 
\end{array}\right)
\right]\,,\\ 
S_{(ij)}(x)&=&S_{ij}\,.
\end{eqnarray}
With their help we can write the action of the spin terms (\ref{sss}) 
and, implicitly, that of the  basis-generators 
$X_{(AB)}^{\rho}=S_{(AB)}^{\rho}+L_{(AB)}$ of the 
representations of $S(M_{\epsilon})$ induced by the representations  
$\rho$ of $SL(2,C)$. 
Hereby it is not difficult to show that  $S(M_{\epsilon})$ is isomorphic 
with the universal covering group of $I(M_{\epsilon})$ which 
in both cases ($\epsilon=\pm1$) is a subgroup of the $SU(2,2)$ group. 
As was expected, in the central charts and Cartesian gauge the 
fields transform manifestly covariant only under the transformations of the 
subgroup $SU(2)\subset S(M_{\epsilon})$.

\subsection{Minkowskian charts}
\

Another possibility is to solve the hyperboloid equation (\ref{hip}) in  
Minkowskian charts \cite{SW} where the coordinates, $x^{\mu}$, are defined 
by 
\begin{equation}
Z^{5}={\hat\omega}^{-1}\tilde\chi_{\epsilon}(s)\,,\quad
Z^{\mu}=x^{\mu}\,,
\end{equation}
with $\tilde\chi_{\epsilon}(s)=\sqrt{1+\epsilon\,\hat\omega^{2}s^{2}}$ 
and $s^{2}=\eta_{\mu\nu}x^{\mu}x^{\nu}$. In these coordinates it is 
convenient 
to identify the hat indices with the usual ones and to do not rise or lower 
these indices. Then we find that the metric tensor, 
\begin{equation}
g_{\mu\nu}(x)=\eta_{\mu\nu}-\frac{\epsilon\,\hat\omega^2}
{\tilde\chi_{\epsilon}(s)^2}
\eta_{\mu\alpha}x^{\alpha}\eta_{\nu\beta}x^{\beta}\,,
\end{equation}  
transforms manifestly covariant under the global $L_{+}^{\uparrow}$ 
transformations, 
$x'^{\mu}\to x^{\mu}=\Lambda^{\mu\,\cdot}_{\cdot\,\nu}x^{\nu}$. Moreover, the 
whole theory remains manifest covariant if we use the tetrad fields in the 
Lorentz gauge defined as \cite{POL}
\begin{equation}
e^{\mu}_{\nu}(x)=\delta^{\mu}_{\nu}+h_{\epsilon}(s)\eta_{\nu\alpha}x^{\alpha}
x^{\mu}\,,\quad 
\hat e^{\mu}_{\nu}(x)=\delta^{\mu}_{\nu}+\hat h_{\epsilon}(s)\eta_{\nu\alpha}
x^{\alpha}x^{\mu}\,, 
\end{equation}
where
\begin{equation}
h_{\epsilon}(s)=\frac{1}{s^2}\left[\tilde\chi_{\epsilon}(s)-1 \right]\,,\quad
\hat h_{\epsilon}(s)=\frac{1}{s^2}\left[\frac{1}{\tilde\chi_{\epsilon}(s)}-1 
\right]\,.
\end{equation}

First we calculate the  $SO(4,1)$ or $SO(3,2)$ orbital generators, 
\begin{eqnarray}
L_{(\mu5)}&=&\frac{i\epsilon}{\hat\omega}\tilde\chi_{\epsilon}(s)
\partial_{\mu}\,,\\
L_{(\mu\nu)}&=&i(\eta_{\mu\alpha}x^{\alpha}\partial_{\nu}
-\eta_{\nu\alpha}x^{\alpha}\partial_{\mu})\,,
\end{eqnarray}
which are independent on the gauge fixing. We observe that in the  
Minkowskian charts 
$\partial_{t}$ is no more a Killing vector as in the case of the central ones. 
However, here we have another advantage namely that of the Lorentz gauge in 
which the local $sl(2,C)$ generators of Eq.(\ref{sss}) have the form
\begin{eqnarray}
S_{(\mu5)}(x)&=&-\frac{\epsilon}{\hat\omega s^2}\left[
\tilde\chi_{\epsilon}(s)-1\right]
S_{\mu\alpha}x^{\alpha}\,,\\
S_{(\mu\nu)}(x)&=&S_{\mu\nu}\,,
\end{eqnarray}
showing that the field $\psi_{\rho}$ transforms manifestly covariant under 
the whole $SL(2,C)$ subgroup of $S(M_{\epsilon})$. Since these representations 
are induced just by those of $SL(2,C)$ we can say that in this gauge the    
manifest covariance is maximal.

\section{Concluding remarks}
\

We have presented here the theory of external symmetry in general relativity.
Starting with the group $I(M)$ which gives the  symmetry of the 
background, we have defined the group $S(M)$ showing that its  Lie algebra, 
$s(M)$, is isomorphic to $i(M)$, having the same structure constants. 
We have pointed out that the fields with spin transform according to the  
representations of the $S(M)$ group induced by linear representations of 
$SL(2,C)$. This allowed us to calculate  the generators of 
these representations  which have specific spin terms. In this way  
we have obtained  the operators associated to the external 
symmetries which commute with the operator of the relativistic  
covariant field equation. We have thus the opportunity to choose suitable 
sets of commuting operators which should determine the quantum modes.

In other respects, since  the concrete form of these generators depends on 
the choice of both the natural and local frames, the commutation rules 
among their spin and orbital parts are determined by the tetrad gauge.  
Consequently,  the results of the local measurement of the spin observables 
may depend on the positions of the local frame. This suggests that it should 
be interesting to investigate new inertial spin effects in other possible 
tetrad gauge fixings.

However, in our opinion, the most important domain of the further 
developmens is that of the external symmetry of the quantum field theory in 
curved spacetimes where the generators of the symmetry transformations 
must be the one-particle operators corresponding to the external symmetries 
through the Noether theorem.\\    
       
{\large \bf Acknowledgments}\\ 

I would like to thank  Mircea Bundaru and  Mihai Visinescu for useful 
comments and enlightening discussions about some sensitive problems 
appearing here.


\begin{thebibliography}{20}

\bibitem{SW}
S. Weinberg, {\it Gravitation and Cosmology: Principles and Applications of 
the General Theory of Relativity}  (Wiley, New York, 1972)


\bibitem{MTW}
C. M. Misner, K. S. Thorne and J. A. Wheeler, {\it Gravitation}  (W. H. Freeman 
\& Co., San Francisco, 1973) 


\bibitem{WALD}
R. M. Wald, {\it General Relativity}, (The Univ. of Chicago Press, Chicago 
and London, 1984)


\bibitem{BD}
N. D. Birrel and P. C. W. Davies, {\it Quantum Fields in Curved Space} 
(Cambridge University Press, Cambridge 1982)


\bibitem{SOL1}
S. J. Avis, C. J. Isham and D. Storey, {\it Phys. Rev. D} {\bf 10}, 3565 (1978);
P. Breitenlohner and D. Z. Freedman, {\it Phys. Lett.} {\bf 115B}, 197 (1982);
D. J. Navarro and J. Navarro-Salas, {\it J. Math. Phys.} {\bf 37}, 6006 (1996); 
I. I. Cot\u aescu, {\em Phys Rev. D} {\bf 60}, 107504 (1999)


\bibitem{SOL2}
V. S. Otchik, {\it Class. Quant. Grav.} {\bf 2}, 539 (1985);
L. P. Chimento and M. S. Mollerach, {\it Phys. Rev. D} {\bf 34}, 3698 (1986);
{\it Phys. Lett. A} {\bf 121}, 7 (1987);
M. A. Costagnino, C. D. El Hasi, F. D. Mozzitelli and J. P. Paz, 
{\it Phys. Lett. A} {\bf 128}, 25 (1988)
V. M. Villalba and U. Percoco, {\it J. Math. Phys.} {\bf 31}, 715 (1990);
G. V. Shishkin, {\it Class. Quant. Grav.} {\bf 8}, 175 (1991);
G. V. Shishkin and V. M. Villalba, {\it J. Math. Phys.} {\bf 30}, 2132 (1989);
{\it J. Math. Phys.} {\bf 33}, 2093 (1992)


\bibitem{UK} 
R. Utiyama, {\em Phys. Rev.} {\bf 101}, 1597 (1956);
T. W. B. Kibble, {\em J. Math. Phys.} {\bf 2}, 212 (1961)
 
\bibitem{W}
W.- K. Tung, {\em Group Theory in Physics}  (World Sci., Philadelphia, 1985)  


\bibitem{ON}
B. O'Neill, {\em Semi-Riemannian Geometry} (Academic Press, 1983)

\bibitem{COTA}
I. I. Cot\u aescu, {\it Mod. Phys. Lett. A} {\bf 13}, 2923 (1998); {\em id} 
{\bf 13}, 2991 (1998); {\em Phys. Rev. D} {\bf 60}, 124006 (1999)  

\bibitem{BJDR} 
J. D. Bjorken and S. D. Drell S.D.  {\it Relativistic Quantum Mechanics} 
(McGraw-Hill Book Co., NY, 1964)

\bibitem{TH} 
B. Thaller,  {\it The Dirac Equation}  (Springer Verlag, Berlin 
Heidelberg, 1992)

\bibitem{BW}
D. R. Brill and J. A. Wheeler, {\em Rev. Mod. Phys.} {\bf 29}, 465 (1957)  

\bibitem{D} 
D. R. Brill and J. A. Cohen, {\em J. Math. Phys.} {\bf 7}, 238 (1966); 
J. Klauder and J. A. Wheeler, {\em Rev. Mod. Phys.} {\bf 29}, 516 (1957); 
T. M. Davis and J. R. Ray, {\em J. Math. Phys.} {\bf 16}, 75 (1975),
{\em Phys. Rev. D}  {\bf 9}, 331 (1974),
J. Math. Phys. {\bf16}, 80 (1975); 
K. D. Kriori and H. Kakati, GRG {\bf 20}, 1237 (1995);
J. C. Huang, N. O. Santos and Kleber, {\em Class. Quantum Grav.} 
{\bf 12}, 1245 (1995);
I. D. Soares and J. Tiomno, {\em Phys. Rev. D} {\bf 54}, 2808 (1996);
C. G. De Oliveira and J. Tiomno, {\em Il Nouvo Cimento} {\bf 24}, 672 (1962);
B. D. B. Figueredo, I. D. Soares and Tiomno, {\em Class. Quantum Grav.} 
{\bf 9}, 1593 (1992);
Hammond R., Class. Quantum Grav. {\bf 12}, 279 (1995); P. Baekler, 
M. Setz, V. Winkelmann, {\em Class. Quantum Grav.} {\bf 5}, 479 (1988);
V. M. Villalba, {\em Mod. Phys. Lett. A}  {\bf 8}, 2351 (1993)


\bibitem{POL}
S. A. Pol'shin, hep-th/0001040; hep-th/0001069

\bibitem{GIL}
R. Gilmore, {\em Lie Groups, Lie Algebras and Some of Their Applications}
(Wiley-Interscience, New York, 1974) 

\bibitem{HAM}
M. Hamermesh, {\it Group theory and its applications to physical problems} 
(Addison-Wesley, Reading MA, 1962)

\bibitem{BR}
A. O. Barut and R. Ra\c czka, {\em Theory of Group Representations and 
Applications}  (PWN, Warszawa, 1977)

\bibitem{MAK}
G. Mackey, {\it Induced Representations of Groups and Quantum Mechanics} 
(Benjamin, New York, 1968)

\bibitem{VIL}
V. M. Villalba, {\em Eur. J. Phys.} {\bf 15}, 191 (1994) 

 
\end{thebibliography}
\end{document}